\documentclass[pra,twocolumn,amssymb,amsmath]{revtex4-1}
\usepackage{CJK}
\usepackage{graphicx}
\usepackage{epsfig,color}
\usepackage{amsmath,amssymb,eufrak}
\usepackage{amsthm}
\usepackage{enumerate}
\usepackage{hyperref}
\usepackage{color,amsxtra,amsfonts,bm}

\def\Id{{\openone}}
\newcommand{\be}{\begin{equation}}
\newcommand{\ee}{\end{equation}}
\newcommand{\bea}{\begin{eqnarray}}
\newcommand{\eea}{\end{eqnarray}}
\begin{document}

\title{Eliminating fermionic matter fields in lattice gauge theories}

\date{\today}

\author{Erez Zohar}
\address{Max-Planck-Institut f\"ur Quantenoptik, Hans-Kopfermann-Stra\ss e 1, 85748 Garching, Germany.}

\author{J. Ignacio Cirac}
\address{Max-Planck-Institut f\"ur Quantenoptik, Hans-Kopfermann-Stra\ss e 1, 85748 Garching, Germany.}

\begin{abstract}
We devise a unitary transformation that replaces the fermionic degrees of freedom of lattice gauge theories by (hard-core) bosonic ones. The resulting theory is local and gauge invariant, with the same symmetry group. The method works in any spatial dimensions and can be directly applied, among others, to the gauge groups $G=U(N)$ and $SU(2N)$, where $N\in\mathbb{N}$. For $SU(2N+1)$ one can also carry out the transformation after introducing an extra idle $\mathbb{Z}_2$ gauge field, so that the resulting symmetry group trivially contains $\mathbb{Z}_2$ as a normal subgroup. Those results have implications in the field of quantum simulations of high-energy physics models.
\end{abstract}

\maketitle

\section{Introduction}

Lattice gauge theories (LGTs) \cite{Wilson,KogutSusskind}
 constitute a powerful way of formulating problems in High Energy Physics. They typically consist of fermionic and bosonic degrees of freedom describing ´the matter and gauge fields, respectively. The different statistical nature of both degrees of freedom is essential to describe most of the fundamental phenomena that occur in nature.

That statistical nature reflects itself when one tries to solve problems in LGTs by means of computers. Fermionic degrees of freedom (or Grassmann variables) cannot be dealt with directly in computations. On the one hand, one has to integrate them out in the path integral formalism, leading to an action that is non-local in the time coordinate. This approach is the basis of Monte-Carlo simulations, the most successful way of solving LGT problems in the non-perturbative regime. For exact diagonalizations,
on the other hand, one also has to convert the fermions to spins (hard-core bosons) to write the wavefunctions in the computational basis.
 This is usually done by means of the Jordan-Wigner transformation \cite{Jordan1928}, which produces non-local models in space (except in 1+1 dimension, where the final theory is local).

In the last few years a growing interest has been built around quantum computers or analog quantum simulators to address LGT problems. In the first case, one also has to represent the fermions in terms of spins (qubits) leading to an overhead in the simulation \cite{Bravyi2002,Jordan2014,Whitfield2016,Bravyi2017}.
 In the second, this restricts the physical setup with which such problems can be addressed. For instance, analog simulators based on ions, superconducting qubits, or photons cannot deal with fermionic degrees of freedom, since the simulator itself is built out of bosons. This is why, beyond one spatial dimension, cold atoms have so far been considered as the best candidate for the task of quantum simulation of dynamical gauge fields with dynamical fermions \cite{Zohar2015a,Wiese2013,Dalmonte2016} (quantum dots, offering fermionic degrees of freedom, could be suitable as well \cite{Barthelemy2013}). Other systems could be used for $1+1d$ - such as superconducting qubits, used, for example, for the quantum simulation of two-dimensional pure-gauge theory \cite{Marcos2014}, or trapped ions, using four of which a quantum simulation of a $U(1)$ lattice gauge theory was carried out in $1+1d$ \cite{Martinez2016}  (see, however, \cite{Lamata2014}).

Some of the above issues may be circumvented if one could map the fermionic degrees of freedom into bosons, while keeping the theory completely local. Apart from raising fundamental questions, like whether fermions are really needed to formulate basic theories, this may have important implications in classical and computational methods to address LGT. In fact, a technique to eliminate the fermionic modes in lattice systems (not necessarily LGT) has already been proposed in \cite{Ball2005,Verstraete2005}. There, auxiliary fermions are introduced in a way that the non-local parts of the Jordan-Wigner strings are cancelled out and, at the same time, a set of local constraints ensure that the spectrum of the Hamiltonian remains invariant.

In this paper we follow a different approach to replace the fermionic by (hard-core) bosonic degrees of freedom in LGTs and identify a unitary transformation that achieves this goal. It rotates the original Hamiltonian in a way that the fermionic modes can be locally transformed into bosonic ones by explicitly using the gauge fields. It requires some auxiliary fermionic degrees of freedom, which we classify into two types, I and II.  Type I are the ones that are finally combined with the original ones to build the bosonic modes. Type II act in a catalytic way; they enable the transformation but remain factorized after it, so that they do not appear in the transformed Hamitonian. Interestingly, the number of degrees of freedom remains invariant: if we initially have $N$ fermionic modes per lattice site, after the transformation we end up with $N$ hard-core bosonic modes as well. Furthermore, the new Hamiltonian is gauge invariant with the same group.
This method can be directly applied to LGT with a gauge group $G$ containing (in its representation) the group element $-\Id$.
This is the case of $G=U(N)$ and $SU(2N)$, in the fundamental representations, commonly used in lattice gauge theories. In order to apply it to the fundamental represenation of $G=SU(2N+1)$, inspired by \cite{Ball2005,Verstraete2005} we first introduce an extra $\mathbb{Z}_2$ gauge field, which is trivially coupled to the fermionic matter, so that then we can apply our approach. In that case, the final theory includes the extra gauge field and is gauge invariant with a larger symmetry group.

This paper is organized as follows: We begin with a review of basic properties of lattice gauge theories in the Hamiltonian formulation, discuss the elimination of fermions for the $U(N)$ and $SU(2N)$ cases and finally the $SU(2N+1)$ case, for which the auxiliary $\mathbb{Z}_2$ gauge field is required.

\section{Lattice gauge theory models} \label{LGTintro}

In this section we will introduce the models that we are going to consider throughout the paper. We will concentrate on 2+1 lattice gauge theories with gauge group $G=U(N)$ or $G=SU(N)$. Models in higher spatial dimensions or with other gauge groups have analogous descriptions. In the first subsection we will express the Hamiltonian describing those models, and in the second one we will briefly review their local symmetries.

\subsection{Hamiltonians}

\begin{figure}
 {\includegraphics[height=40em]{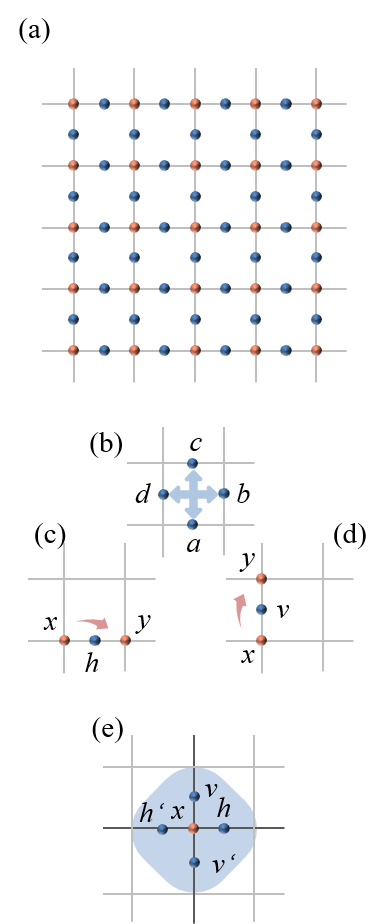}}
 \caption{General lattice gauge theory settings. (a) The fermionic fields reside on vertices, while the gauge fields on the links of a square lattice.
 Notation conventions for: (b) the plaquette interaction of $H_{\text{B}}$ as used in (\ref{HB}); (c),(d) the link interactions of $H_{\text{I}}$ (\ref{Hhorvert}); (e) Gauss law (\ref{Gauss}).} \label{fig1}
\end{figure}

We consider a 2+1 dimensional lattice gauge theory in the Hamiltonian formulation. In a square lattice (see Fig. \ref{fig1}a), fermionic modes are associated to the vertices (red circles), whereas gauge fields
 to the links (blue circles). Denoting by ${\cal H}_{\rm m}$ and ${\cal H}_{\rm fields}$ the Hilbert spaces corresponding to the fermions and the bosons, respectively, the physical Hilbert space, ${\cal H}_{\rm phys}$, is a subspace of ${\cal H}_{\rm m}\otimes {\cal H}_{\rm fields}$, that is determined by the Gauss law (see below).

We assume there are $N$ fermionic modes at each vertex $x$, with annihilation operators $\psi_{x,m}$, ($m=1,\ldots,N$) fulfilling the  canonical anti-commutation relations, and   collected in a spinor $\psi_x$. The gauge fields at each link $\ell$ are described in terms of a set of operators, $L_{\ell,\alpha},R_{\ell,\alpha}$, and a $N\times N$ matrix of operators, $U_\ell$, which commute at different links. At the same link, they fulfill
 \begin{subequations}
 \label{LaRa}
 \bea
 \label{La}
 \big{[}L_\alpha,U_{n,m}\big{]} &=& \sum_{n'} (T_\alpha)_{m,n'} U_{n',n},\\
 \big{[}R_\alpha,U_{n,m}\big{]} &=& \sum_{n'} U_{m,n'}(T_\alpha)_{n',n}, \eea
 \end{subequations}
where we have omitted the link index, $\ell$ for the sake of clarity. We will do that in the following wherever it does not lead to confusion.
The matrices $T_\alpha$ are the $N$-dimensional representations of the gauge group generators.
As mentioned in the introduction, here we consider the groups $G=U(N)$ or $G=SU(N)$
in the fundamental representation (which are used in most physical contexts and are $N$ dimensional).

The Hamiltonian can be written as
 \be
 \label{Hdef}
 H = H_{\text{M}} + H_{\text{E}}+ H_{\text{B}} + H_{\text{I}}
 \ee
Here, $H_{\text{M}}$ is the local (mass) part of the matter field,
 \be
 \label{HM}
 H_{\text{M}} = M \sum_{x,m} (-1)^{\left|x\right|} \psi_{x}^\dagger \psi_{x}
 \ee
where according to the notation introduced above
 \be
 \psi_x^\dagger \psi_x = \sum_{m=1}^N \psi_{x,m}^\dagger \psi_{x,m} =: n_x.
 \label{ndef}
 \ee
(where we have chosen, for convenience, the staggered fermionic formulation \cite{Susskind1977,Zohar2015}; other formulations of lattice fermions could be used as well).
The electric ($H_{\text{E}}$) and magnetic ($H_{\text{B}}$) parts of the Hamiltonian describe the gauge fields, and are given by \cite{KogutSusskind}
 \begin{subequations}
 \bea
 H_{\text{E}} &=& \lambda_E \sum_\ell J_\ell^2,\\
 H_{\text{B}} &=& \lambda_B \sum_{\mathfrak{p}} {\rm Tr}\left(U_a U_b U_c^\dagger U_d^\dagger\right) + h.c.  \label{HB}
 \eea
 \end{subequations}
Here, $J^2=\sum_\alpha L_\alpha^2 = \sum_\alpha R_\alpha^2$ is the Casimir operator at each link, $\mathfrak{p}$ runs over all plaquettes in the lattice, and for each plaquette $a,b,c,d$ are chosen in the order shown in Fig. \ref{fig1}b. Finally, the gauge-fields-matter interaction is given by $H_{\text{I}}=H_{\rm hor}+ H_{\rm vert}$, where
  \begin{subequations}
 \label{Hhorvert}
 \bea
 H_{\rm hor} &=& \epsilon \sum_h \psi^\dagger_x U_h \psi_y + h.c., \\
 H_{\rm vert} &=& \epsilon \sum_v \psi^\dagger_x U_v \psi_y + h.c.
 \eea
 \end{subequations}
contain the hopping of matter along the horizontal (Fig. \ref{fig1}c) and vertical (Fig \ref{fig1}d) links, with the sum extended to all the horizontal (vertical) links, and where $x,y$ are the left (down) and right (up) vertices with respect to $h$ ($v$). We have carried out the separation between horizontal and vertical interactions since they will have a slightly different role in the transformation that we will carry out in the next sections.

\subsection{Gauge invariance}

The Hamiltonians appearing in (\ref{Hdef}) are all invariant under local unitary transformations acting around each vertex, $x$. The physical space, ${\cal H}_{\rm phys}$, must be an eigenspace of such operators, which is defined according to the Gauss law in terms of the corresponding generators
 \be
 \label{Gauss}
 \left(G_{x,\alpha} - Q_{x,\alpha}\right)|\Psi\rangle=0,
 \ee
for any $\Psi\in{\cal H}_{\text{phys}}$, $x$ and $\alpha$, where
 \begin{subequations}
 \bea
 G_{x,\alpha} &=& L_{h,\alpha}+L_{v,\alpha}-R_{h',\alpha}-R_{v',\alpha},\\
 \label{Charge}
 Q_{x,\alpha} &=& \psi^\dagger_x T_\alpha \psi_x
 \eea
 \end{subequations}
and the labeling of the links around $x$ is defined in Fig. \ref{fig1}e (the definition of $Q_{x,\alpha}$ for $U(N)$ is slightly different and takes different forms on the two sublattices corresponding to particles and anti-particles in the staggered formulation \cite{Susskind1977,Zohar2015}, but this has no relevance to this work). The states fulfilling (\ref{Gauss}) span the physical space.

\subsection{Examples: U(1) and SU(2) lattice gauge theories}
\label{subExamples}

Let us first give the simplest example, where $G=U(1)$ \cite{KogutLattice}.
In that case: (i) $G$ is abelian, so that we can omit the index $\alpha$ in (\ref{LaRa}) since it only takes one value. We have $L=R=:E$, which plays the role of the electric field; (ii) $G_x=E_{h}+E_{v}-E_{h'}-E_{v'}$ is just the (discrete) divergence of the electric field at position $x$; (iii) $N=1$, so that we have just one fermionic mode per lattice site, and $U$ is just a unitary operator fulfilling $\left[E,U\right]=U$. This automatically implies that
 \be
 \label{PU}
 PU+UP=0
 \ee
where $P=e^{i\pi E}$. Furthermore,
 \be
 \label{PE}
 [P,E]=0
 \ee
follows trivially.

Let us now to move to the case $G=SU(2)$: (i) $G$ is non-abelian, and there are three generators ($\alpha=x,y,z$); (ii) since $N=2$, there are two fermionic modes at each site, and we can take in (\ref{LaRa}) $2T_\alpha=\sigma_\alpha$, the Pauli matrices; (iii) From (\ref{La}) it immediately follows that
 \be
 e^{i \theta L_\alpha} U e^{-i \theta L_\alpha}=e^{i \theta \sigma_\alpha/2} U.
 \ee
Defining $P=e^{i 2\pi i L_z}$
we obtain again (\ref{PU}) since $\exp(i\pi \sigma_\alpha)=-\Id$ for any Pauli matrix.
This can be easily obtained by defining $f_\alpha(q)=\exp(i q L_z) L_\alpha \exp(-i q L_z)$, taking the derivatives with respect to $q$, and solving the resulting differential equations with the help of
 \be
 \left[L_\alpha,L_\beta\right]=-i \varepsilon_{\alpha,\beta,\gamma} L_\gamma,
 \ee
where $\varepsilon_{\alpha,\beta,\gamma}$ is the (Levi-Civita) completely antisymmetric tensor.
Thus, again, there exist (in fact, an infinite number of) operators $P$ fulfilling (\ref{PU}) (since we can choose any direction in the Bloch sphere for $\alpha$, or take $R_\alpha$ instead of $L_\alpha$). Furthermore,
 \be
 \left[P,L_\alpha\right]=\left[P,R_\alpha\right]=0
 \ee
for any $\alpha$, which is the analog of (\ref{PE}).  Note that for $G=SU(2)$, in order to fulfill (\ref{PU}), we must use a half-integer representation (i.e., the $T_\alpha$ are $M\times M$ matrices with $M$ even).

The existence of an operator $P$ fulfilling (\ref{PU}) and (\ref{PE}) will be the basis of the method we introduce in the following in order to transform the fermions in the model into hard cord bosons. As we will explain in the next section, this occurs for $U(N)$, as well as for $SU(2N)$ as long as we deal with the fundamental representation, as we are considering here.

\section{Elimination of the fermionic degrees of freedom: $U(N)$ and $SU(2N)$}

The objective of this section is to introduce some auxiliary degrees of freedom and a unitary transformation which allows us to map the fermionic modes into bosonic ones. In the first subsection we will describe the basic idea behind the method. In the second we will introduce a set of auxiliary fermionic degrees of freedom, and the unitary operator. In the last one we will show how the Hamiltonian is transformed under such an operation, leading to a hard-core boson lattice gauge theory.

\subsection{Method}
\label{idea}

Let us briefly explain the main idea of the method. The detailed explanation will be given in the rest of this section. We introduce in each vertex two kinds of auxiliary fermionic modes, which we will call type I and II. The new physical Hilbert space is ${\cal H}_{\text{phys}}\otimes \Omega_I \otimes \Omega_{II}$, where  $\Omega_O$ ($O=I,II$) is a one dimensional space containing the vacuum space of the auxiliary modes, $|\Omega_O\rangle$. The Hamiltonian is the one introduced in the previous section, $H$ (\ref{Hdef}), which acts trivially on the auxiliary modes, so that this model is completely equivalent to the original one.

We will define a unitary transformation, ${\cal U}$, and rotate the Hilbert space and the Hamiltonian with it. This entangles the auxiliary modes with the original ones and the gauge fields. An important property of ${\cal U}$ is that this only occurs locally. We will denote ${\cal H}'_{{\rm phys}}={\cal U} \left({\cal H}_{\rm phys} \otimes \Omega_I \otimes \Omega_{II} \right)$ and $H'={\cal U} (H\otimes \Id_{I} \otimes \Id_{II}) {\cal U}^\dagger$. It turns out that the transformation leaves  $\Omega_{II}$ invariant, so that we can write ${\cal H}'_{\rm phys}=\tilde {\cal H}_{{\rm phys}}\otimes \Omega_{II}$. Thus, we can restrict the transformed Hamiltonian to this space by taking $\tilde H= \langle \Omega_{II}|H'|\Omega_{II}\rangle$, so that we are effectively left with the original modes and the auxiliary ones of type I only.

The two main features of the new Hamiltonian are: (i) each fermionic operator $\psi_{x,m}$ will be replaced by $\eta_{x,m}:=c_x \psi_{x,m}$, where $c_x$ is a majorana operator of the type I auxiliary fermions defined at vertex $x$; (ii) the operators $U$ at each link will be replaced by $U$ times a gauge field operator involving the fields surrounding the vertices attached to that link. The first one (i) already implies that we are left with bosonic operators, since the operators $\eta_{x}$ commute at different sites,
$\left[\eta_x,\eta_y\right]=\left[\eta_x,\eta_y^\dagger\right]=0$ if $x\ne y$. Furthermore, since $\eta_m^2=0$, we can associate them to the creation of hard-core bosons. In fact, we can define a local Hilbert space at each vertex, ${\cal H}_x$ which is generated by the action of the $\eta_x^\dagger$ on the vacuum (for the original and type I fermions), and
 \be
 \tilde {\cal H}_{\rm m}= \oplus_x {\cal H}_x
 \ee
The physical space in the rotated frame, spanned by the rotated states $\left|\tilde{\Psi}\right\rangle = \left\langle \Omega_{II}\right|\mathcal{U}\left|\Psi\right\rangle\left|\Omega_I\right\rangle\left|\Omega_{II}\right\rangle$,  will be $\mathcal{H}_{\text{phys}} \subset \tilde{\cal H}_{\rm m}\otimes\tilde{\cal H}_{\rm fields}$, defined by the new Gauss law, which will be the same as in (\ref{Gauss}) but with the trivial replacement of the fermionic charges by the hard-core ones, i.e.
 \be
 \label{Qtransf}
  Q_{x,\alpha}\to \tilde Q_{x,\alpha} = \eta^\dagger_x T_\alpha \eta_x.
 \ee
The second feature of the transformation, (ii), will modify the form in which the gauge fields enter $H_{\rm hor}$,$H_{\rm vert}$ (\ref{Hhorvert}), and $H_{\text{B}}$ (\ref{HB}), by adding some gauge field dependent signs. Those will still be local, but will break the 90-degree rotational invariance of the Hamiltonian. They take care of the signs that are left by the replacement of the fermions by the bosons.

The spectra of $H$ and $\tilde H$ are identical, so that if one is just interested in the spectral properties of $H$, one can equally work with the latter. The physical properties can be also be computed with it via the Hellmann-Feynman theorem \cite{Feynman1939}, by applying the unitary transformation to the extra terms that are introduced for that purpose.
In order to compute expectation values of physical (gauge invariant) observables, $A$, one could first compute the state in the original picture $|\Psi\rangle=\langle\Omega_{II},\Omega_I|{\cal U}^\dagger(|\tilde\Psi\rangle\rangle|\Omega_{II}\rangle$ and from it the expectation value. One would be tempted to ignore $|\Omega_{II}\rangle$ in this computation, since it is invariant. However, this is not possible since the unitary transformation involves fermions of type II, it will introduce non-trivial phases in the wavefunction $|\Psi\rangle$. Therefore, it is more convenient to compute all the expectation values of observables directly in the transformed picture, i.e. to determine $\tilde A:=\langle \Omega_{II}|{\cal U} A {\cal U}^\dagger|\Omega_{II}\rangle$, which will keep the locality of observables. In this section we will also show how to compute such transformations.

\subsection{Auxiliary fermions and the transformation}

In Section \ref{subExamples} we showed that for $G=U(1)$ and $G=SU(2)$ it is always possible to find an operator $E$ such that for $P=\exp\left({i\pi E}\right)$ we have (\ref{PU},\ref{PE}). In fact, this construction is always possible for $G=U(N)$ and $G=SU(2N)$, as  shown in App. \ref{appa} (there we also include other groups). Thus, since we are considering these cases in this section, we will assume that such an operator $E$ exists in each link.

\begin{figure}
 {\includegraphics[height=30em]{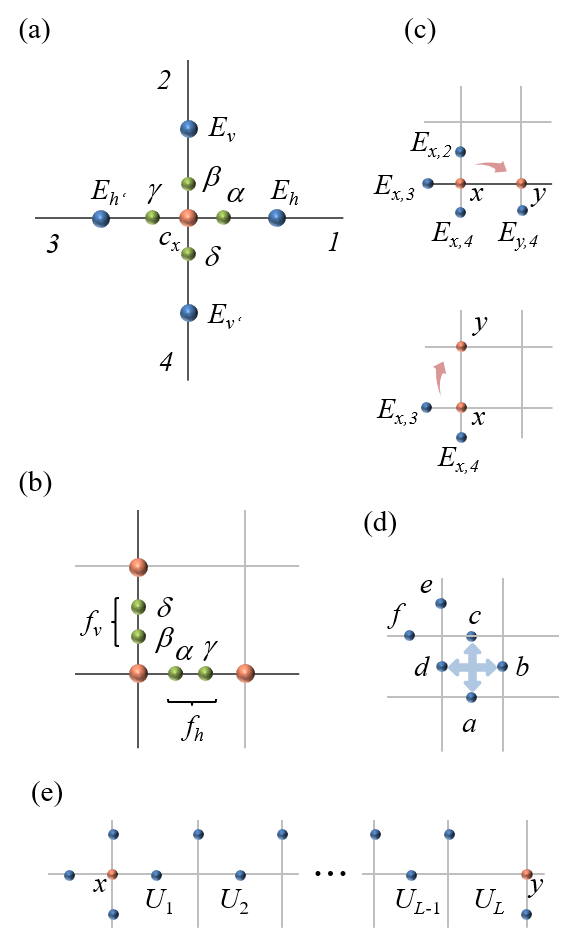}}
 \caption{Notation conventions for the transformation and its results. (a) Auxiliary fermions of I - $c_x$, and of type II - $\alpha_x,\beta_x,\gamma_x,\delta_x$ and $E$ conventions around a single vertex $x$, for (\ref{VVdef}); (b) The fermionic modes along links  $f_{h,v}$ are built out of the Majorana modes associated with the link's endpoints, (c) the sign factors $\xi_{h,v}$ (\ref{sgnfactors}); (d) sign factors for $\tilde{H}_{\text{B}}$ (\ref{HBtr}) ; (e) sign factors $\xi_{\mathcal{M}}$ for the transformed horizontal meson operator $\tilde{\mathcal{M}}$ (\ref{mestr}) - links with circles contribute to the sign factor.} \label{fig2}
\end{figure}

At each vertex $x$, we define now a fermionic mode of type I, with annihilation operator $\chi_x$, and a majorana operator $c_x=\chi_x+\chi_x^\dagger$. We also introduce four majorana operators for the fermionic modes of type II, $\alpha,\beta,\gamma,\delta$, as  shown in Fig. \ref{fig2}a. We associate each of them to one link around the vertex $x$, according to the drawing (Fig. \ref{fig2}b). Thus, for each link we will have two such majorana operators, each of them defined on a contiguous vertex. As usual, all the majorana operators anticommute with each other and square to the identity operator. With those operators we build a fermionc annihilation operator on each horizontal, $h$, and vertical, $v$, link, as indicated in the figure:
 \be
 f_h = \frac{1}{2}(\alpha_h-i\gamma_h),\quad f_v=\frac{1}{2}(\beta_v-i\delta_v).
 \ee
We also define the vacuum states, $|\Omega_I\rangle$, $|\Omega_{II}\rangle$, as the ones annihilated by those operators:
 \begin{subequations}
 \bea
 \chi_x |\Omega_I\rangle &=&0,\\
 f_h|\Omega_{II}\rangle &=& f_v|\Omega_{II}\rangle=0.
 \eea
 \end{subequations}

Now we build the unitary transformation. It is defined as
 \be
 {\cal U} = \prod_x {\cal U}_x
 \ee
where ${\cal U}_x$ is built out the fermionic operators (of type I and II) on the vertex $x$, as well as the operators $E$ corresponding to the links connected to that vertex. Specifically,
 \be
 \label{Ulocdef}
 {\cal U}_x = V_{x,4} V_{x,3} V_{x,2} V_{x,1}
 \ee
where
 \begin{eqnarray*}
 V_1 &=& (i c \alpha)^{E_h}, \quad V_2= (i c \beta)^{E_v},\\
 V_3 &=& (i c \gamma)^{E_{h'}}, \quad V_4= (i c \delta)^{E_{v'}},\\
 \label{VVdef}
 \end{eqnarray*}
where the numbering of the operators $V_i$ follows Fig. \ref{fig2}a.

The first thing to notice is that the operator ${\cal U}$ is unitary. This follows from the fact that the operators $V_i$ are unitary (note that $(i c \alpha)^2=\Id$ and  $V_i^\dagger=V_i$). Furthermore, $[{\cal U}_x,{\cal U}_y]=0$ since: (i) regarding the gauge degrees of freedom, they only depend only on the operators $E$, which commute among themselves; (ii) the fermionic operators belonging to each vertex only appear at that vertex and are even in creation and annihilation operators. This implies that the operator ${\cal U}$ does not depend on the order in which we multiply the ${\cal U}_x$. However, the operators $V_i$ at a given vertex do not commute among themselves, so that the transformed Hamiltonian will depend on the order in which have chosen their product in ${\cal U}_x$.

\subsection{The Transformation}

In order to determine the transformed Hamiltonian under ${\cal U}$, we just need to know how the different operators appearing in it transform individually. Since the operators $\psi$ and $L_\alpha$ and $R_\alpha$ commute with ${\cal U}$, we just need to find out how $U$ transforms. We have to distinguish between the horizontal and vertical links (see App. \ref{appb}):
 \be
 {\cal U} U_r {\cal U}^\dagger = i \xi_r \left(1-2f_rf^\dagger_r \right) c_x  U_r c_y,
 \label{Urtrans}
 \ee
where $r=h,v$, and $x,y$ are the two vertices connected to $r$ (see Fig. \ref{fig1}c,d). The phase factors $\xi$ are given by (see Fig. \ref{fig2}c)
 \begin{subequations}
 \label{sgnfactors}
 \bea
 \xi_h &=& e^{i\pi (E_{x,2}+E_{x,3}+E_{x,4}+E_{y,4})},\\
 \xi_v &=& e^{i\pi (E_{x,3}+E_{x,4})}
 \eea
 \end{subequations}
The inclusion of those factors allows us to maintain the commutation relations between the transformed $U$'s at neighboring sites, as should happen for unitary transformations.

The emergence of the operators of type II is only in the form $f_r f_r^\dagger$. All those commute among themselves, and thus they also do with the transformed Hamiltonian. In fact, in App. \ref{appb} it is shown that
 \be
 {\cal U} f_r f_r^\dagger {\cal U}^\dagger = f_r f_r^\dagger
 \ee
This has two consequences: (i) $f_r {\cal U} ({\cal H}_{\rm phys}\otimes |\Omega_I\rangle\otimes|\Omega_{II}\rangle)=0$, and thus in the transformed space the state of the type II fermions is still the vacuum; (ii) we can replace $f_rf_r^\dagger\to 1$ in the transformed Hamiltonian if we project it onto that vacuum, as anticipated in Section \ref{idea}.

As advanced, we will define
 \be
 \eta_m = c \psi_m
 \label{etadef}
 \ee
at each lattice site. On the same site, we have
 \be
 \psi^\dagger_m \psi_n = \eta_m^\dagger \eta_n
 \label{psieta}
 \ee

For the $G=U(1)$ case we can replace $\eta,\eta^\dagger$ with Pauli operators $\sigma_{\pm}$. For $N>1$ we can still define the $\eta_m$ in terms of such operators for $N$ spins, through a process that involves a local Jordan-Wigner transformation (see App. \ref{appb}).

\subsection{Transformed Hamiltonian}

The transformed Hamiltonian, once projected onto $\Omega_{II}$, can be written as $\tilde H= \tilde H_{\text{M}} + \tilde H_{\text{E}} + \tilde H_{\text{B}} + \tilde H_{\text{I}}$. The first two are unaffected by the transformation, $\tilde H_{\text{E}}=H_{\text{E}}$ and [compare Eq. (\ref{HM})]
 \be
 \tilde H_{\text{M}} = M \sum_{x,m} (-1)^{\left|x\right|} \eta_{x}^\dagger \eta_{x}
 \ee

The interaction Hamiltonian can be written as a sum over the horizontal and vertical links (\ref{Hhorvert}). According to the discussion of the previous subsection, they are transformed as
 \begin{subequations}
 \bea
 \tilde H_{\rm hor} &=& -i \epsilon \sum_h \xi_h \eta^\dagger_x U_h \eta_y + h.c., \\
 \tilde H_{\rm vert} &=& -i \epsilon \sum_v \xi_v \eta^\dagger_x U_v \eta_y + h.c.
 \eea
 \end{subequations}

Finally, the magnetic part results in (see App. \ref{appb})
 \be
 \tilde H_{\text{B}} =  \lambda_B \sum_{\mathfrak{p}} \xi_{\mathfrak{p}} {\rm Tr}\left(U_a U_b U_c^\dagger U_d^\dagger\right)  + h.c.
 \label{HBtr}
 \ee
where
 \be
 \xi_{\mathfrak{p}}= e^{i\pi (E_a+E_b+E_e+E_f)}
 \ee
and the links $a,b,e,f$ are the ones shown in Fig. \ref{fig2}d. The transformation of other gauge invariant observables are given in App. \ref{appb}.

The new Hamiltonian $\tilde H$ is gauge invariant with the same group $G$ as the original Hamiltonian $H$. The Gauss law still holds (\ref{Gauss}) but with the substitution (\ref{Qtransf}). Relevant physical quantities in this transformed picture will be the transformed operators $\tilde A=\langle \Omega_{II}|{\cal U} A {\cal U}^\dagger|\Omega_{II}\rangle$ defined in Sec. \ref{idea}. These consists of electric field operators (invariant under $\mathcal{U}$), Wilson loops and mesonic operators, that transform similarly to the plaquette and link interactions - which are the smallest Wilson loops or mesons. Thus, for example, the horizontal meson operator
\begin{equation}
\mathcal{M}=\psi^{\dagger}_x U_1 U_2 ... U_{L} \psi_{y}
\label{mes0}
\end{equation}
where $x,y$ are two points on the same horizontal line, such that $y$ is located $L$ links to the right of $x$, and $U_1, U_2, ..., U_{L}$ are defined on these $L$ links, will be transformed to
\begin{equation}
\tilde{\mathcal{M}} = -i^L\xi_{\mathcal{M}}\eta^{\dagger}_x U_1 U_2 ... U_{L} \eta_{y}
\label{mestr}
\end{equation}
 where
$\xi_{\mathcal{M}} = \left(-1\right)^{\underset{i}{\sum}E_i}$ and $E_i$ that are summed are defined in Fig. \ref{fig2}e.

\subsection{Discussion: $\mathcal{U}$, Statistics and the Unitary Gauge}

After having seen the action of the unitary transformation $\mathcal{U}$, let us explain the physical grounds of that procedure and describe the different roles of the auxiliary fermions of type I and II.

$\mathcal{U}$ could be seen as a fermionic version of the \emph{unitary gauge} of
   the \emph{Brout-Englert-Higgs mechanism} \cite{Brout1964,Higgs1964,Fradkin1979}, \footnote{Note that we build an analogy with the gauge transformation that takes place in that mechanism, and not about the symmetry breaking it is related to there.}.
Let us consider, as an example, a case similar to that discussed in \cite{Fradkin1979}, where a $U(1)$ gauge field on the links ($U_{\ell}=e^{i \phi_{\ell}}$, $E_{\ell}$ as in Sec. \ref{subExamples}) is coupled to a complex scalar (bosonic) matter field residing on the vertices, represented by
\begin{equation}
\Phi_{x} = e^{-i \theta_x}\rho_{x}
\label{scalpol}
\end{equation}
Note that $\Phi_{x}$ is a complex scalar field operator, and not a mode operator: it contains two bosonic degrees of freedom, and therefore $\left[\theta_x,\rho_{x}\right]=0$.
In the unitary gauge fixing procedure carried out in \cite{Fradkin1979}, the $U(1)$ phase (Goldstone mode) $\theta_x$ is absorbed by the gauge field,
i.e. $\phi_{\ell} \rightarrow \phi_{\ell} + \theta_y - \theta_x$ (where $x$ ($y$) is the beginning (end) of the link $\ell$.) by a transformation which is a product of the local unitaries
\begin{equation}
\mathcal{U}^{\text{scalar}}_x = e^{i \left(E_h + E_v - E_{h'} - E_{v'}\right) \theta_x}
\label{Uscalar}
\end{equation}
defined at each vertex, using the notation introduced in the current work (see Fig. \ref{fig2}a).

In our case, we have fermions instead of bosons. One can then invert the definition of $\eta_{x,m}$ (\ref{etadef}) to a form analogous to the scalar one (\ref{scalpol}),
\begin{equation}
\psi_{x,m}=c_x\eta_{x,m} = e^{- i \pi \left(c_x-1\right)/2}\eta_{x,m}
\label{ferpol}
\end{equation}
 Instead of the bosonic $U(1)$ phase operator $e^{-i\theta_x}$, here we have a fermionic "$\mathbb{Z}_2$ phase operator", $c_x$. The analogy of the $U(1)$ phase $\theta_x$ is the $\mathbb{Z}_2$ "phase" $\pi \left(c_x-1\right)/2$ (note that, as in \cite{Fradkin1979}, one could also have $\mathbb{Z}_N$ gauge fields, including $\mathbb{Z}_2$, coupling to Higgs matter of the same group, but it is always bosonic, even in the $\mathbb{Z}_2$ case, unlike here).  One could thus be tempted to eliminate the fermions with a transformation analogous to (\ref{Uscalar}), i.e.
 \begin{equation}
 \mathcal{U}^{\text{fer}}_x = e^{i\pi \left(E_h + E_v - E_{h'} - E_{v'}\right)\left(1-c_x\right)/2}
  \end{equation}
  The problem is that this transformation does not preserve the fermionic parity; it can be either even or odd, depending on the divergence $E_h + E_v - E_{h'} - E_{v'}$. Therefore, in general different $\mathcal{U}^{\text{fer}}_x$ transformations on different vertices do not commute, and for their product some order must be chosen.

This problem is solved by introducing the auxiliary fermions of type II; in fact, the local transformation $\mathcal{U}_x$ (\ref{Ulocdef}) simply satisfies
 \begin{equation}
 \mathcal{U}_x \propto \mathcal{U}^{\text{fer}}_x
 \end{equation}
 the remaining pieces that we did not write here explicitly (and could be obtained by bringing (\ref{Ulocdef}) to the form above by re-ordering the fermionic operators) depend on the $E$ operators and the type II auxiliary modes. They complete $\mathcal{U}^{\text{fer}}_x$ to an even $\mathcal{U}_x$, that preserves the parity locally and allows one to perform the transformation safely. Indeed, the type II modes have no physical role and they only account for balancing the fermionic parity, and therefore they can be easily factored out eventually.

Unlike in the bosonic case,  here $\left\{c_x,\eta_{x,m}\right\}=0$. In fact, $c$ is an extra degree of freedom; no symmetry is broken here, and no degree of freedom is eliminated by the absorption of $c$ by $U$ in the transformation. On the contrary, the physical Hilbert space after the transformations seems to have been enlarged, since now it includes these type I fermions in a non-trivial way. However, as shown in App. \ref{appb}, the transformed state obeys the relation
\begin{equation}
e^{i \pi \underset{m}{\sum}\eta^{\dagger}_{x,m}\eta_{x,m}}\left|\tilde{\Psi}\right\rangle
=e^{i\pi \chi^{\dagger}_{x}\chi_{x}}\left|\tilde{\Psi}\right\rangle
\label{statphys}
\end{equation}
for every vertex $x$. It manifests a local $\mathbb{Z}_2$ symmetry, that connects the \emph{statistics}, having to do with the local fermionic parity - now carried by the auxiliary fermions of type I - with the parity of the total number of physical excitations, which are not fermionic in the transformed picture (in App. \ref{appb} we show how to solve this constraint).
The above discussion implies that the method requires that the gauge group $G$ includes $\mathbb{Z}_2$ as a normal subgroup (this is not a sufficient condition; one also needs to use a representation that includes $-\Id$); indeed, in App. \ref{appa} we will show that such gauge groups allow for a construction of a $P$ operator, for particular representations.

\section{Elimination of the fermionic degrees of freedom: other cases}

The fermion elimination procedure described above applies neither to  gauge groups of the form $SU(2N+1)$ nor to others that, for instance, do not possess a normal $\mathbb{Z}_2$ subgroup. For such groups, $P,E$ operators satisfying (\ref{PU}) cannot be defined (see App. \ref{appa}). In this section, we present a method that is valid for any group. The main idea is to couple the fermions to an auxiliary gauge field such that: (i) the operators $P,E$ exist; (ii) the model possesses an additional gauge invariance; (iii) by including appropriate Gauss laws, the new Hamiltonian is equivalent to the original one. Once this is done, the method used in the previous section can be directly applied to eliminate the fermionic degrees of freedom.

The method can also be applied to local fermionic models with nearest neighbor hopping that do not include gauge fields. In fact, it shares certain analogies with those introduced in Refs. \cite{Ball2005,Verstraete2005} for those purely fermionic models, and where auxiliary $\mathbb{Z}_2$ fields with constraints were also considered. In both cases, the physical theory is embedded in another that includes auxiliary gauge fields and an extra local $\mathbb{Z}_2$ symmetry, while maintaining the physical properties of the original model. There are, however, some significant differences. On the one hand, in \cite{Ball2005,Verstraete2005} the auxiliary gauge field was constructed out of fermionic objects on the links that fused with the original fermions to form hard-core bosons. In our case, the auxiliary fermions that fuse to hard-core bosons (type I) reside on the vertices, and emerge from a unitary transformation that involves the local gauge field, as explained in the previous section. This enables to fuse the physical fermions with the auxiliary ones in a way that depends only on the vertices and not the links, without having to impose additional constraints as in \cite{Ball2005}, for example. On the other hand, as we will see, we impose an extra Gauss law that includes the fermions and the auxiliary fields to make the new Hamiltonian unitarily equivalent to the original one.

\subsection{Introducing Auxiliary $\mathbb{Z}_2$ Gauge Fields} \label{MZ2}

\begin{figure}
 {\includegraphics[height=30em]{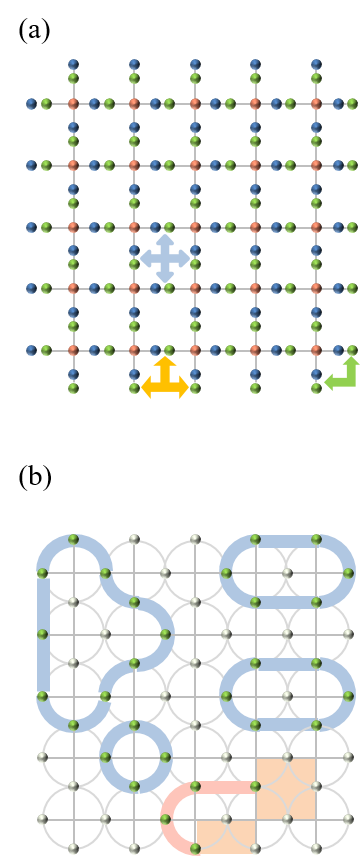}}
 \caption{(a) System with open boundary conditions. Apart from the original gauge field degrees of freedom (blue dots), there are auxiliary gauge degrees of freedom (green) on the links; the three kinds of plaquettes are indicated: bulk (blue), boundary (orange), and corner (green); (b) Loop configuration. Highlighted are the spins in $\left|1\right\rangle$. All blue loops are closed and thus they respect that Gauss law (\ref{Gauss3}), whereas the one in red is open, and thus violates that law at the edges, i.e. in the plaquettes marked in red.} \label{fig3}
\end{figure}

So far we have not specified what type of boundary conditions we are considering since this was not required. In most parts of the present section we will consider open boundary conditions (OBC) as shown in Fig. \ref{fig3}a. That is, the lattice ends up with links in which gauge fields are defined. Later on, we will discuss how to extend the discussion to periodic boundary conditions (PBC).

In order to proceed, on each link we add an additional bosonic  gauge field (see Fig. \ref{fig3}a) that takes the form of a qubit, with the orthonormal basis $|0\rangle$ and $|1\rangle$. We will denote by $X=\sigma_x$ and $Z=\sigma_z$ the Pauli operators acting on the qubit, with the convention $Z|0\rangle=|0\rangle$ and $X|0\rangle=|1\rangle$. We define
 \begin{subequations}
 \label{Hhorvert2}
 \bea
 H_{\rm hor}' &=& \epsilon \sum_h \psi^\dagger_x  U_h' \psi_y + h.c., \\
 H_{\rm vert}'&=& \epsilon \sum_v \psi^\dagger_x U_v' \psi_y + h.c.
 \eea
 \end{subequations}
where $U'_\ell =Z_\ell U_\ell$ for every link $\ell=h,v$. The new Hamiltonian takes the form $H'=H_{\text{M}}+H_{\text{E}}+H_{\text{B}}+H_{\rm hor}^{\rm aux}+H_{\rm ver}^{\rm aux}$.
It has many local symmetries, so that we define the new physical space, ${\cal H}_{\rm phys}'$ as that obeying the following Gauss laws: (i) the Gauss law corresponding to the original Hamiltonian, i.e. (\ref{Gauss}); (ii) an analogous one, related to the auxiliary gauge field:
 \be
 \label{Gauss2}
 A_x e^{i\pi n_x}|\Psi\rangle = |\Psi\rangle
 \ee
where
 \be
 \label{Ax}
 A_x=X_hX_vX_{h'}X_{v'}
 \ee
and $n_x$ is defined in (\ref{ndef}); (iii) finally, one associated with to the auxiliary gauge field alone (Gauss law on the dual lattice). In each plaquette $\mathfrak{p}$ we impose
 \be
 \label{Gauss3}
  \prod_{\ell\in \mathfrak{p}}  Z_\ell |\Psi\rangle = |\Psi\rangle,
 \ee
There are three kind of plaquettes: at the bulk, at the boundary, and at the corners. Each of them gives rise to the product of four, three, and two $X$ operators, respectively (see Fig. \ref{fig3}a). Note that $Z_\ell^2=\Id$ and $(A_xe^{i \pi n_x})^2=\Id$, so that the complete gauge group is $G\times \mathbb{Z}_2\times \mathbb{Z}_2$ (the third is on the dual lattice). We will denote the new physical Hilbert space by ${\cal H}_{\rm phys}'$.

With these definitions, we have that: (i) as we will show in the next subsection, the Hamiltonian $H'$ restricted to the physical space ${\cal H}_{\rm phys}'$ is unitarily equivalent to the original one, $H$, restricted to its corresponding one, ${\cal H}_{\rm phys}$; (ii) The operator $P'=X=e^{i\pi E'}$, with $E'=(1-X)/2$ fulfills (\ref{PU}) and (\ref{PE}), i.e., $PU'+U'P'=0$ and $\left[P',E'\right]=0$.
Therefore, we can directly apply the procedure of the previous section to obtain a model where the fermions are replaced by hard-core bosons, with an extra gauge field (a qubit per link), and the new Gauss laws (\ref{Gauss2}) and (\ref{Gauss3}). While (\ref{Gauss2}) remains the same (with (\ref{psieta})), the other Gauss law constraint (\ref{Gauss3}) becomes, at the bulk plaquettes
 \be
 Y_aY_b Z_cZ_d X_e X_f|\tilde\Psi\rangle = -|\tilde\Psi\rangle
 \ee
 with the labels of Fig. \ref{fig2}d,  where $Y=\sigma_y$, and similarly at the other ones (the boundary and the corners). This is proven in the same way as the transformation rule of the magnetic Hamiltonian in the previous case, as given in App. \ref{appb}.

\subsection{$H$ and $H'$ are unitarily equivalent on the physical space}

We now first characterize the space ${\cal H}_{\rm phys}'$ and show that it is isomorphic to ${\cal H}_{\rm phys}$. Then, we will show that the Hamiltonians $H'$ and $H$, restricted to those spaces, are related by a unitary transformation which we will identify.

Let us denote by ${\cal H}_{\rm aux}$ the space of all qubit configurations fulfilling the second Gauss law (\ref{Gauss3}). That space is characterized by the fact that in each plaquette there is an even number of qubits in state $|1\rangle$, as it has been extensively studied in the literature \cite{Levin2005}. This is most easily characterized if we draw curves that join pairs of qubits in that state at each plaquette (see Fig. \ref{fig3}b). In that case, the only allowed curves are closed curves, i.e. loops \cite{Levin2005,Kitaev2003}, and we can thus establish an orthonormal basis in ${\cal H}_{\rm aux}$, $\{|L\rangle\}_{L\in \Lambda}$, where $\Lambda$ is the set of all possible loop patterns. We will call $|L=0\rangle$ that which corresponds to no loop (i.e. all qubits in $|0\rangle$). Notice that each state $|L\rangle$ can be created out of $|L=0\rangle$ by acting with the operators $A_x$ (\ref{Ax}) on specific vertices $x$. This can be seen as follows: let us take a state with an arbitrary loop pattern and pick a particular loop. If we now apply $A_x$ to all the vertices enclosed within that loop, we will obtain a state with the same loop pattern as the original one, but without that loop. In this way, we can get rid of all the loops sequentially and end up with the state $|L=0\rangle$ for any $|L\rangle$. We denote by $A_L$  the operator that maps $A_L|L\rangle=|L=0\rangle$. Since $A_L^2=\Id$, $A_L|L=0\rangle=|L\rangle$, i.e. one can generate any state in the orthonormal basis out of the state without loops. The total number of such operators is $2^{N_v}$, where $N_v$ is the number of vertices in the lattice, since for each vertex, $x$, we can either apply $A_x$  or not. Accordingly, the dimension of ${\cal H}_{\rm aux}$ is $2^{N_v}$.

Obviously, ${\cal H}_{\rm phys}'\subset {\cal H}_{\rm phys}\otimes {\cal H}_{\rm aux}$. Thus, for any state $|\Psi'\rangle \in {\cal H}_{\rm phys}'$ we can write
 \be
 \label{PsiL'}
 |\Psi'\rangle = \frac{1}{2^{N_v/2}}\sum_{L\in{\cal L}} |\Phi_L\rangle\otimes |L\rangle
 \ee
where $|\Phi_L\rangle\in {\cal H}_{\rm phys}$. The state $|\Psi'\rangle$ trivially fulfills the Gauss laws (\ref{Gauss}) and (\ref{Gauss3}), so that we just have to impose (\ref{Gauss2}). For any loop pattern, $L\in\Lambda$, we define
 \be
 n_L=\sum_{x\in L} n_x
 \ee
and the sum is extended to all the vertices that are inside the loops contained in $L$. A state fulfilling (\ref{Gauss2}) must be invariant under $A_L e^{i\pi n_L}$ for all $L\in \Lambda$. This immediately implies
 \be
 |\Phi_L\rangle = e^{i \pi n_L} |\Phi_{L=0}\rangle
 \ee
so that
 \be
 |\Psi'\rangle = \frac{1}{2^{N_v/2}}\sum_{L\in{\cal L}} e^{i \pi n_L}|\Psi\rangle\otimes|L\rangle.
 \ee
This establishes an isomorphism $W: {\cal H}_{\rm phys}\to {\cal H}_{\rm phys}'$, with $W:|\Psi\rangle\to |\Psi'\rangle$. Furthermore,
 \begin{subequations}
 \bea
 \langle \Psi'_1|\Psi'_2\rangle &=& \langle \Psi_1|W^\dagger W |\Psi_2\rangle = \langle \Psi_1|\Psi_2\rangle,\\
 \langle \Psi'_1|H'|\Psi'_2\rangle &=&\langle \Psi_1|W^\dagger H' W |\Psi_2\rangle = \langle \Psi_1|H|\Psi_2\rangle
 \eea
 \end{subequations}
so that, as claimed, $H$ and $H'$ are unitarily equivalent in their physical spaces. Moreover, expectation values of $\mathbb{Z}_2$ gauge-invariant objects (e.g., fermionic operators connected with $U'$ operators) have the same expectation values in ${\cal H}_{\rm phys}'$ as the analogous ones, but now connected with $U$ operators in the original space ${\cal H}_{\rm phys}$. This implies that the extended theory is completely equivalent to the original one, so that we can work with $H'$ instead of $H$.

Let us finish this section with a comment about periodic boundary conditions. In that case, the space ${\cal H}_{\rm aux}$ of all loops is not fully generated by the $A_L$. One needs to define $A_{\rm hor}$ ($A_{\rm vert}$) which are built by taking a product of $X_v$ ($X_h$) along a horizontal (vertical) line wrapping the torus. The whole space has four sectors, corresponding to acting with $A_{\rm hor}^{n_h}A_{\rm ver}^{n_v}$ on the space generated by the $A_L$, with $n_h,n_v=0,1$. They, in fact, correspond to the four topological sectors that appear in the Toric code \cite{Kitaev2003}. The Hamiltonian $H'$ is only unitarily equivalent to $H$ only in the sector with $n_h=n_v=0$.

\section{Conclusion}

In this work, we have shown how fermions in lattice gauge theories whose gauge group is $U(N)$ or $SU(N)$ may be mapped to hard-core bosonic degrees of freedom in a local way. This is based on the local $\mathbb{Z}_2$ symmetry.

The method presented above is valid, in general, for lattice gauge theories with dynamical fermions, as well as to pure fermionic Hamiltonians for which there exists a well-defined minimal coupling procedure. It suggests a way to explore fermionic theories circumventing the computational difficulties of anti-commutativity or fermion-based non-locality, as everything is done in a local manner. While the paper was written for $SU(N)$ and $U(N)$ lattice gauge theories in $2+1$ dimensions, it can be generalized in a straightforward manner to other dimensions (including, in particular, $1+1$ and $3+1$, or any other dimension), as well as other gauge groups.

\begin{acknowledgments}
 EZ would like to thank  D. Gonz\'alez Cuadra, A. Moln\'ar and A. Sterdyniak for helpful discussions. JIC is partially supported by the EU, ERC grant QUENOCOBA 742102.
 \end{acknowledgments}

\appendix

\section{Construction of $P$ Operators} \label{appa}

In this appendix, we will give the explicit construction of the operator $P$ satisfying (\ref{PU}) for $U(N)$ and $SU(2N)$, through its relation to the $\mathbb{Z}_2$ subsymmetry.

The relation (\ref{PU}) may be rewritten as $PUP = -U$, which means that the operator $P$ is a group transformation that multiplies the matrix $U$ by $-\Id$. As $-\Id$ commutes with any matrix, it is not important if $P$ is realized as a left or as a right transformation, i.e. generated using $L_{\alpha}$ or $R_{\alpha}$. We conclude that in order for such an operator to exist, there must be a group element $z \in G$, which, in the irreducible representation we work with, is represented by $-\Id$. This is true for $U(N)$ in any representation, and in particular the fundamental one that we would like to use. For $SU(N)$, in the fundamental representation whose dimension is $N$, it is true only for an even $N$.

This group element $z$ squares to the identity, and forms, with the identity element $e$, a normal $\mathbb{Z}_2$ subgroup of $G$. Thus, we conclude that an operator $P$ satisfying (\ref{PU}) only exists (and not necessarily uniquely) for groups $G$ with a normal $\mathbb{Z}_2$ subgroup, in representations for which it is represented as $-\Id$.

In the $U(1)$ case we saw how to construct such an operator. In the $U(N)$ case, since $U(N)=SU(N) \times U(1)$, the $U(1)$ which is not contained in $SU(N)$ forms a normal subgroup of $U(N)$ (as its elements commute with all the rest of the group's elements), and thus one can use the abelian electric field $E$ belonging to that component and construct $P=e^{i\pi E}$ out of it as in the $U(1)$ case.

In the case of $SU(2N)$, $-\Id_{2N\times 2N}$ is the fundamental representation of the $z \in SU(2N)$ we need. In the main text, its construction for $SU(2)$ was discussed, and it can be extended to any even $N$ using the generator whose fundamental representation is
\begin{equation}
T_{0} = \frac{1}{\sqrt{2N\left(N-1\right)}}\left(
              \begin{array}{cccccc}
                1 &   &   &   &  & \\
                  & 1 &   &   & &  \\
                  &   & 1 &   &  & \\
                  &   &  & \ddots  &  & \\
                  &   &   & &1  &\\
                  &   &   &   & &-(N-1) \\
              \end{array}
            \right)
\end{equation}
(for example, in $SU(2)$ this is $\sigma_z/2$).
Then, and only for an even $N$,
\begin{equation}
e^{i \pi \sqrt{2N\left(2N-1\right)} T_0} = -\Id
\end{equation}
and one can use either the left generator $L_0$ corresponding to it (or the right one $R_0$), to define and construct $P$ and $E$:
\begin{equation}
P = e^{i \pi \sqrt{2N\left(2N-1\right)} L_0} \equiv e^{i \pi E}
\end{equation}
giving us the desired result (\ref{PU}) when $U$ is in the fundamental representation. This is not the case for any representation; for example, as discussed in the main text, in $SU(2)$ this applies only for representations with an even dimension (half-integer spins).

The same can be done for the finite groups $\mathbb{Z}_{2N}$ as well.
In any $\mathbb{Z}_N$ model, the link Hilbert space is $N$ dimensional. In it, one defines two unitary operators $Z,X$, satisfying \cite{Horn1979}
\begin{equation}
\begin{aligned}
Z^N=X^N=\Id \\
XZX^{\dagger} = e^{2\pi i /N}Z
\label{ZN}
\end{aligned}
\end{equation}
The spectrum of both operators is the set of $N$th roots of unity, $\left\{e^{2\pi i m /N}\right\}_{m=0}^{N-1}$.
In $\mathbb{Z}_N$ lattice gauge theories, $Z$ takes the role of the $U$ operator.

For $N=2$, we simply get that $P=X$, and, in general, for $\mathbb{Z}_{2N}$, agreeing with the main text. In general, for $\mathbb{Z}_{2N}$, $P = X^{N/2}$,
as it is responsible, through the $\mathbb{Z}_{N}$ relations (\ref{ZN}), to putting the phase $ e^{\left(2\pi i /N\right) \times \left(N/2\right)} = -\Id$ on $Z$ and hence (\ref{PU}) is satisfied by it.

\section{Properties of $\mathcal{U}$} \label{appb}

In this appendix we will discuss some general properties of the transformation $\mathcal{U}$ and derive the transformation rules of some operators used in the main text.

\subsection{The Transformation's Building Blocks}

The local transformation $\mathcal{U}_x$ (\ref{Ulocdef}) is constructed out of the ingredients $V_i=\left(ic\zeta_i\right)^{E_i}$ (\ref{VVdef}), where $\zeta_i = \alpha,\beta,\gamma,\delta$ is a type II auxiliary majorana mode at the vertex associated with some direction, and $E_i$ is the $E$ operator on the link emanating from the vertex to that direction. We introduce $C_i = i c \zeta_i$, such that $V_i = C_i^{E_i}$.
Since $C_i^2 = \Id$, it can be expressed as
\begin{equation}
C_i = ie^{-i \pi C_i/2}=e^{i \pi \left(1-C_i\right)/2}
\label{ZZZ}
\end{equation}
as well as
\begin{equation}
V_i = C_i^{E_i} = e^{i \pi E_i \left(1-C_i\right)/2} = P_i^{\left(1-C_i\right)/2}
\label{VVV}
\end{equation}
As the spectrum of $C_i$ is $\pm 1$, that of $\left(1- C_i\right)/2$ is $0,1$.

Using (\ref{PU}), (\ref{ZZZ}) and (\ref{VVV}) we can  derive the transformation properties of $U$ under a single $V_i$ operation -
\begin{equation}
V_i U V_i^{\dagger} = C_i U = i c \zeta_i U
\label{VUV}
\end{equation}
- if $V_i$ belongs to one of the edges of the link on which $U$ resides. Otherwise they commute.

 Another relevant transformation is
\begin{equation}
V_i c V_i^{\dagger} = \zeta_i^{E_i} c^{2E_i+1} \zeta_i^{E_i} =e^{i \pi E_i}c = P_i c
\label{ctrans}
\end{equation}
(if $V_i$ and $c$ belong to the same vertex; otherwise they commute).

Combining the two, we obtain the transformation of $U$. For a $U$ operator on the link connecting $x,y$, only the local transformations $\mathcal{U}_x,\mathcal{U}_y$ contribute,
and their order is not important since they commute. In each of them, the first $V_i$ that will not commute through $U$ is the one having to do with the link's direction as in (\ref{VUV}), adding, in particular, a $c$ operator that does not commute with the remaining $V_i$ operators that will act according to (\ref{ctrans}). As this depends on the ordering of $V_i$ operators in $\mathcal{U}_x$, the results for horizontal and vertical links will be different, giving rise to (\ref{Urtrans}) with the sign factors $\xi_r$.

The $U$ operators on different links commute, and this commutation should be preserved by any unitary transformation, in particular $\mathcal{U}$. This is possible thanks to the $\xi_r$ factors. Had they not appeared, the transformed $U$ operators on intersecting links, due to the fermionic operators that are added to them by the transformation, would have anti-commuted after the transformation. The additional $\xi_r$ factors, introducing $P$ operators on some neighboring links, convert this anti-commutation to commutation again, thanks to the anticommutativity of (\ref{PU}).

\subsection{Transformation of the Auxiliary Fermions}

The auxiliary fermions of type II transform similarly to the $c$ modes (\ref{ctrans}):
\begin{equation}
\mathcal{U} \zeta_i \mathcal{U}^{\dagger} = c^{E_i} \zeta_i^{2E_i+1} c^{E_i} = e^{i \pi E_i}\zeta_i = P_i \zeta_i
\end{equation}
Combining the two type II modes associated with one particular link - for example, $\alpha$,$\gamma$ when it is horizontal, we obtain the invariance:
\begin{equation}
\mathcal{U} f_h f_h^{\dagger} \mathcal{U}^{\dagger} = \frac{1}{2} \mathcal{U} \left(1+i\alpha_h\gamma_h\right) \mathcal{U}^{\dagger} = \frac{1}{2}\left(1+i\alpha_h\gamma_h\right) = f_h f_h^{\dagger}
\end{equation}
(using the notation conventions of Fig. \ref{fig2}b). Similarly, $\mathcal{U} f_v f_v^{\dagger} \mathcal{U}^{\dagger} = f_v f_v^{\dagger}$ for vertical links.

Finally, let us see how the auxiliary vertex modes transform:
\begin{equation}
\begin{aligned}
\mathcal{U}\chi^{\dagger}_{x}\chi_{x}\mathcal{U}^{\dagger} &=
c_{x}^{\left(E_h + E_v - E_{h'} - E_{v'}\right)}\chi^{\dagger}_{x}\chi_{x}c_{x}^{\left(E_h + E_v - E_{h'} - E_{v'}\right)} \\&=
\frac{1}{2} \left(1-e^{i\pi \left( \left(E_h + E_v - E_{h'} - E_{v'}\right) - \chi^{\dagger}_{x}\chi_{x}\right)}\right)
\end{aligned}
\end{equation}
Thus, the initial condition $\chi^{\dagger}_{x}\chi_{x}\left|\Psi\right\rangle = 0$ transforms to
$e^{i\pi  \left(E_h + E_v - E_{h'} - E_{v'}\right)}\left|\tilde{\Psi}\right\rangle = e^{i\pi  \chi^{\dagger}_{x}\chi_{x}}\left|\tilde{\Psi}\right\rangle$; combining that with the gauge invariance (\ref{Gauss}) - or, in particular, its $\mathbb{Z}_2$ normal subgroup, we obtain the relation (\ref{statphys}) - a constraint that reduces the size of the enlarged Hilbert space, that now, instead of only the physical fermions, involves the type I auxiliary ones as well. Later in this appendix we will show how this constraint could be removed (along with the auxiliary degrees of freedom of type I) using a local unitary transformation on top of local Jordan Wigner transformations.

\subsection{Transformation of the Magnetic Hamiltonian}

The remaining transformation left to perform is this of the magnetic Hamiltonian $H_{\text{B}}$. Using (\ref{Urtrans}), one may transform the single plaquette operator as
\begin{widetext}
\begin{equation}
\begin{aligned}
\left\langle \Omega_{II} \right| \mathcal{U}
\text{Tr}\left(
U_1
U_2
U^{\dagger}_3
U^{\dagger}_4
\right)
\mathcal{U}^{\dagger}
\left| \Omega_{II} \right\rangle&=
\left\langle \Omega_{II}\right|
i^4\left(1-2f_af_a^{\dagger}\right)\left(1-2f_bf_b^{\dagger}\right)\left(1-2f_cf_c^{\dagger}\right)\left(1-2f_df_d^{\dagger}\right) \left| \Omega_{II}\right\rangle
c_{da}^2c_{ab}^2c_{bc}^2c_{cd}^2
 \\ &\times
\underset{m,n,n',m'}{\sum}
\xi_a\left(U_a\right)_{mn}
\xi_b\left(U_b\right)_{nn'}
\xi_c\left(U^{\dagger}_c\right)_{n'm'}
\xi_d\left(U^{\dagger}_d\right)_{m'm}
\end{aligned}
\end{equation}
\end{widetext}
where the plaquette's links are labeled by $a,b,c,d$ as in Fig. \ref{fig1}b, and $c_{ij}$ is the type I auxiliary fermion at the intersection of the $i,j$ vertices.

The contribution of auxiliary fermions is trivial. The $c$ modes are all squared and give rise to 1, and
$\left\langle \Omega_{II}\right|
i^4\left(1-2f_af_a^{\dagger}\right)\left(1-2f_bf_b^{\dagger}\right)\left(1-2f_cf_c^{\dagger}\right)\left(1-2f_df_d^{\dagger}\right) \left| \Omega_{II}\right\rangle =1$
 as well. One has to simply collect all the $\xi_r$ operators (some do not commute with the $U$ operators on their sides), to obtain $\tilde{H}_{\text{B}}$ (\ref{HBtr}).

\subsection{Representing the Hard-Core Bosons by Spins}

The transformation $\mathcal{U}$ allows one to replace the fermions by hard-core bosonic operators, $\eta_{x,m}$ (\ref{etadef}), $N$ such operators per vertex. Here we shall show how these could be represented by $N$ spin operators.

Let us concentrate on a vertex, and define a local Jordan-Wigner transformation \cite{Jordan1928} (since the Hamiltonian and all physical operators we discuss have a local even fermionic parity, non-local strings do not have to be attached). This requires to numerate the fermionic modes in the vertex and replace them by spins: we take the logical order assigning $m$ to the fermionic mode associated to $\psi_m$ and $N+1$ to that of $c$. Thus we have
\begin{equation}
\begin{aligned}
\psi_{m} &\leftrightarrow i\sigma^{-}_{m} \Sigma_m \Sigma_1\\
c_x &\leftrightarrow \sigma^x_{N+1} {\Sigma}_1 \\\end{aligned}
\end{equation}
where $\sigma$ are Pauli operators,
 \be
 \Sigma_m= \sigma^z_N \sigma^Z_{N-1}\ldots \sigma^z_m,
 \ee
so that
 \be
 \eta_m \leftrightarrow c\psi_m = -i \sigma^x_{N+1} {\Sigma}_{m+1} \sigma^-_m
 \ee
with $\Sigma_{N+1}=\Id$. Now, applying the unitary transformation
 \be
 \tilde W = e^{i \pi \sigma^x_{N+1} \sum_{m=1}^N \sigma^z_m/4}
 \ee
we obtain
 \be
 \tilde \eta_m = \tilde W \eta_m \tilde W^\dagger \leftrightarrow {\Sigma}_{m+1} \sigma^-_m
 \ee
so that the auxiliary spin does not appear anymore and thus can be ignored.

Note that what we have carried out just amounts to replacing $N+1$ fermions by $N$ spins, a common procedure given the fermion parity superselection rule. That is, the states generated by $\eta_m^\dagger$ out of the fermionic vacuum (in our case, annihilated by $\psi_m$ and $\chi$) span the subspace with an even number of fermions, and thus has dimension $2^N$. The transformation above replaces the auxiliary spin by the one that accounts for the parity, which can thus be ignored. Note further that the fermionic vaccum is transformed to the state with all spins in $0$.

\bibliography{ref}

\end{document}